\documentclass {JHEP3}

\usepackage {amsmath}
\usepackage {epsfig}

\newcommand{\bea}{\begin{eqnarray}}
\newcommand{\eea}{\end{eqnarray}}

\newcommand {\be} {\begin{equation}}
\newcommand {\ee} {\end{equation}}

\newcommand{\vol}{\mbox{vol}}
\newcommand{\es}[2] {\begin{equation} \label{#1} \begin{split} #2 \end{split} \end{equation}}
\newcommand{\abs}[1]{\lvert #1 \rvert}

\title {Emergent Quantum Near-Criticality\\ from Baryonic Black Branes}

\author {Christopher P.~Herzog$^*$, Igor R.~Klebanov$^{*\dagger}$, Silviu S.~Pufu$^*$, and Tiberiu Tesileanu$^*$ \\
$*$ Joseph Henry Laboratories, Princeton University, Princeton, NJ 08544, USA  \\
$\dagger$ Princeton Center for Theoretical Science, Princeton University, Princeton, NJ 08544, USA \\
}

\preprint {PUPT-2318}

\abstract{
We find new black 3-brane solutions describing the ``conifold gauge theory'' at nonzero temperature and baryonic chemical potential. Of particular interest is the low-temperature limit where we find a new kind of weakly curved near-horizon geometry; it is a warped product $AdS_2 \times {\mathbb R}^3\times T^{1,1}$ with warp factors that are powers of the logarithm of the AdS radius. Thus, our solution encodes a new type of emergent quantum near-criticality.  We carry out some stability checks for our solutions. We also set up a consistent ansatz for baryonic black 2-branes of M-theory that are asymptotic to $AdS_4\times Q^{1,1,1}$.
}

\begin{document}


\section{Introduction}

Via the AdS/CFT correspondence \cite{Maldacena:1997re,Gubser:1998bc,Witten:1998qj},
electrically charged black holes in space-times with negative cosmological constant
yield insights into the physics of strongly-interacting systems at nonzero density.
Recall that the correspondence relates semi-classical gravity in $d+1$ space-time dimensions to a strongly-interacting field theory in one fewer dimensions.  There have been two kinds of approaches to these problems.  In the bottom-up approach, a simple and phenomenological gravity model is constructed to which the AdS/CFT dictionary is then applied. It is assumed that there exists some field theory, of which we may have only a qualitative understanding, dual to this space-time, but using the correspondence, we can quickly and efficiently determine the phase structure, the equation of state, and transport coefficients.  In the top-down approach, one considers a well-established AdS/CFT duality where the field theory is well-known but may have exotic symmetries and field content.  Given the complexity of the known dual pairs, calculations are often more difficult. However, they are worth the extra effort, because they allow us to make precise and reliable statements about actual strongly interacting field theories.

In this paper, we take the second approach and construct a novel type of charged black hole (or, more precisely, black 3-brane) using the baryonic symmetry of the conifold gauge theory \cite{Klebanov:1998hh} dual to the $AdS_5\times T^{1,1}$ background of type IIB string theory. Our solution is similar to the Reissner-Nordstr\" om $AdS_5$ black hole but is more complicated because the compact space $T^{1,1}$ gets squashed by functions that depend on the radius. In the zero-temperature limit, we find that the near-horizon region becomes similar to $AdS_2 \times {\mathbb R}^3\times T^{1,1}$ up to slowly varying logarithmic functions. The presence of the logarithms makes our IR solution a new kind of nearly conformal behavior. Thus, very interestingly, our solution exhibits ``emergent quantum near-criticality,'' which could make it useful for exploring connections with condensed matter phenomena.

Recently, there has been much interest in bottom-up approaches to study field theories that undergo superfluid or superconducting phase transitions
(see \cite{Hartnoll:2009sz, Herzog:2009xv} for reviews). Strong electron-electron interactions are believed to play an important role in the physics of high-temperature superconductors. Refs.\ \cite{Lee:2008xf,Liu:2009dm, Cubrovic:2009ye,Faulkner:2009wj} use a bottom-up approach to model a strongly-interacting system of fermions at nonzero density, and find evidence for the existence of a Fermi surface.

While the bottom-up approaches allow one to scan quickly through a number of simple gravity models and search for new phenomena, they have some disadvantages.  A major issue is that the precise nature of the dual field theory is typically unclear, and one cannot be certain that it exists.  Another disadvantage is related to the notion of a consistent truncation and its stability.  The AdS/CFT correspondence in its original incarnation is a mapping between type IIB string theory in an $AdS_5 \times S^5$ background and ${\mathcal N}=4$ super Yang-Mills theory in 3+1 dimensions.  In order to reduce a ten-dimensional string theory to a manageable five-dimensional gravity theory, a consistent truncation is made that eliminates all but a small number of fields. The consistency of the truncation guarantees that a solution to the five-dimensional equations of motion for the remaining fields is also a solution to the full ten-dimensional system.  However, nothing guarantees that this solution is a global or local minimum of the action in the ten-dimensional setting;  indeed, often it is not. The simple gravity models in these bottom-up approaches would be consistent truncations if fit into the larger AdS/CFT framework, and as such they may have instabilities.

In top-down approaches, one of the easiest ways to charge a black hole is through the R-symmetry.  The dual field theory often has an R-symmetry which maps to a gauge field in the gravity system.  In the grand canonical ensemble, R-charge on the black hole translates into nonzero R-charge chemical potential $\mu$ in the field theory.  In the most studied case of ${\mathcal N}=4$ super Yang-Mills in 3+1 dimensions, it is strongly suspected that at any nonzero $\mu$, the theory is only metastable \cite{Yamada:2006rx,Yamada:2008em,Yamada:2007gb}. Moreover, if $\mu$ becomes sufficiently large compared to the temperature, the theory becomes thermodynamically and perturbatively unstable as well \cite{Cai:1998ji,Chamblin:1999tk,Cvetic:1999ne}.

More generically, charged scalars, if their charge-to-mass ratio is sufficiently large, are a source of instabilities.  Starting with refs.\ \cite{Gubser:2008px,Hartnoll:2008vx}, these charged scalars have been studied intensively in the context of modeling a superfluid or superconducting phase transition.  Truncating such a charged scalar out of the gravity model also spuriously eliminates the phase transition. While it is true that the physical systems of interest often have a superconducting phase transition at very low temperatures, from a theoretical point of view, it is of value to have a model where one may reliably go to low temperatures without worrying about such instabilities.

In this paper, we construct a black 3-brane charged under a baryonic $U(1)_B$ symmetry that we hope is immune from the instabilities that plague R-charge black holes near extremality.  We consider the well-known conifold gauge theory and its dual pair, type IIB string theory in an $AdS_5 \times T^{1,1}$ background \cite{Klebanov:1998hh}. The $SU(N) \times SU(N)$ field theory with bifundamental fields $A_i$ and $B_j$, $i, j=1,2$, has a global baryonic $U(1)_B$ symmetry. The corresponding $U(1)_B$ gauge field in $AdS_5$ comes from the R-R 4-form with 3 indices along a topologically non-trivial 3-cycle. This realization of the $U(1)$ symmetry makes our approach different from the previous attempts to embed charged AdS black holes into string theory. In particular, the nature of the charged objects is quite different.  The gauge invariant operators with baryonic charge in the conifold gauge theory have conformal dimensions of order $N$.  The smallest such operator in the conifold theory involves determinants of the bifundamental matter fields.  In the string dual, such an operator maps to a wrapped D3-brane which may be studied semi-classically.  We are able to show explicitly that this wrapped D3-brane has a charge-to-mass ratio that is too small to produce an instability. This check, however, is insufficient to demonstrate the stability of our solution because one of the neutral fields may condense as the temperature is decreased. We demonstrate stability with respect to one seemingly dangerous neutral mode, but leave investigation of other modes for the future. We proceed with the optimistic assumption that our solutions are stable even at $T=0$.

In section \ref{sec:setup}, we review the details of the conifold gauge theory and its gravity dual.  We also demonstrate a consistent truncation of type IIB supergravity (SUGRA) for the conifold background to a baryonic gauge field and two neutral scalars in five dimensions.  Given the effective 5d Lagrangian, in section \ref{sec:blackholes} we construct a metric, scalar, and gauge field ansatz for a baryonically charged black 3-brane that depends only on a single radial coordinate.  The ansatz is invariant under a certain $\mathbb{Z}_2$ symmetry and leads to a system of non-linear ordinary differential equations.  In section \ref{sec:numerics}, we find their numerical solutions with $AdS_5\times T^{1,1}$ boundary conditions at large $r$. Although we encounter difficulties at very low temperatures, the numerical work provides us with useful intuition concerning the $T=0$ solution.
In section \ref{sec:zeroT} we find the near-horizon series expansion in the $T=0$ limit and show that it has a near-$AdS_2$ structure.
In section \ref{sec:noPT}, we study the behavior of the smallest operator with baryonic charge in the conifold theory, the dibaryon.  We show that its charge-to-mass ratio is too small to lead to an instability.
In section \ref{sec:stability} we carry out another stability check of our $T=0$ solution, this time with respect to a neutral $\mathbb{Z}_2$-odd perturbation.  Because of nontrivial mixing with the $U(1)_R$ gauge field, the analysis requires us to generalize the ansatz of Section \ref{sec:setup}.  Section \ref{sec:discussion} contains some final remarks and discussion.
In Appendix \ref{sec:smallcharge}, we find an analytic expression for our black hole in the small charge limit.  In Appendix \ref{sec:Qbranes} we consider black 2-branes carrying baryonic charges in M-theory on $AdS_4\times Q^{1,1,1}$, and develop a consistent ansatz and the equations satisfied by these solutions.


\section{Conifold Gauge Theory and Consistent Truncation}
\label{sec:setup}

The principal aim of this paper is to study the AdS/CFT duality in presence of a chemical potential for baryon number. In a general large $N$ gauge theory, the operators carrying baryon number have dimensions of order $N$, which distinguishes them from the
``mesonic'' operators whose dimensions are of order 1 in the large $N$ limit. In AdS/CFT, the objects dual to baryonic operators are D-branes or M-branes wrapped over non-trivial cycles of the internal manifold. In the maximally supersymmetric version, which relates the ${\cal N}=4$ SYM theory to $AdS_5\times S^5$, there are no baryonic operators.
Their absence is related to the fact that $S^5$ has no topologically non-trivial 3-cycles that could be wrapped by D3-branes.
However, there are many known examples where the AdS/CFT duality relates $AdS_5\times Y$, where $Y$ is a Sasaki-Einstein manifold, to ${\cal N}=1$ superconformal gauge theories that have baryonic operators. The compact space $Y$ typically has non-trivial 3-cycles, so that the wrapped
D3-branes are topologically stable and the corresponding gauge theory possesses baryonic $U(1)_B$ symmetries.
We would like to turn on a chemical potential for the baryon number in the gauge theory; in the string dual this translates into
turning on the R-R 4-form gauge field that couples to the wrapped D3-branes.

We will present perhaps the simplest example of such a construction, in the context of the duality relating the $AdS_5\times T^{1,1}$ background of type IIB string theory to the ${\cal N}=1$ superconformal ``conifold gauge theory'' \cite{Klebanov:1998hh}. This duality is motivated by studying a stack of $N$ D3-branes
placed at the tip of the conifold, the Calabi-Yau cone $zw-uv=0$. The explicit metric of its Sasaki-Einstein base, the $T^{1,1}$, is
\es{T11Metric}{
ds_{T^{1,1}}^2 = {1\over 6} \sum_{i=1}^2 \left( d \theta_i^2 + \sin^2 \theta_i d\phi_i^2 \right)
+ {1\over 9} \left(d\psi + \cos \theta_1 d\phi_1
+ \cos \theta_2 d\phi_2 \right)^2 \ .
}
The conifold gauge theory with gauge group $SU(N)\times SU(N)$ is coupled to bi-fundamental chiral superfields $A_i$, $i=1,2$ transforming as
$(\bf{N}, \bf{\bar N})$, and $B_j$, $j=1,2$ transforming as $(\bf{\bar N}, \bf{N})$. These superfields form doublets under the global symmetries $SU(2)_A$ and $SU(2)_B$, respectively, and all of them carry R-charge $1/2$.
In addition to the $U(1)_R \times SU(2)_A \times SU(2)_B$ global symmetry, which on the gravity side is realized through isometries of $T^{1,1}$, the gauge theory has a baryonic $U(1)_B$ symmetry under which
\begin{equation}
A_k \to e^{i \theta} A_k\ , \qquad B_l \to e^{-i\theta} B_l \ .
\end{equation}
The spectrum of gauge invariant operators splits into two sectors. The mesonic operators, which are not charged under the $U(1)_B$, have dimensions of order $1$; the baryonic operators, which are charged under the $U(1)_B$, have dimensions of order $N$.
The lowest dimension examples of mesonic operators are ${\rm Tr} A_i B_j$ of dimension $3/2$, and
${\rm Tr} A_i\bar A_j$,  ${\rm Tr} B_i\bar B_j$, ${\rm Tr} (A_i\bar A_i- B_j\bar B_j)$ of dimension 2.
In general, the mesonic operators transform under the $U(1)_R \times SU(2) \times SU(2)$ geometric symmetry of $T^{1,1}$, but
are neutral under the $U(1)_B$. The lowest dimension operators carrying the $U(1)_B$ charge are, for example,
${\rm det} A_1$ or  ${\rm det} A_2$. These are the $m=\pm N/2$ states of the spin $N/2$ representation of $SU(2)_A$. The general form of these dimension $3N/4$ operators, ${\cal A}_m$, $m=-N, -N+1, \ldots, N-1, N$, may be found in \cite{Gubser:1998fp}. They carry R-charge $N/2$, and we will normalize their baryon number to $1$. The string theory objects dual to these operators are the D3-branes wrapping the $(\theta_1,\phi_1,\psi)$ directions. Quantization of the $(\theta_2,\phi_2)$ collective coordinate gives rise to the $N+1$ degenerate ground states corresponding to the chiral operators ${\cal A}_m$.

Similarly, there exist chiral operators ${\cal B}_m$, $m=-N, -N+1, \ldots, N-1, N$, such that
${\cal B}_N= {\rm det} B_1$ and ${\cal B}_{-N}= {\rm det} B_2$. These operators have baryon number $-1$ and R-charge $N/2$; they are dual to D3-branes wrapping the $(\theta_2,\phi_2,\psi)$ directions. Replacing these D3-branes by anti-D3 branes we find objects of baryon number $1$ and R-charge $-N/2$ that are dual to the antichiral operators ${\cal \bar B}_m$ that include ${\rm det} \bar B_1$.
Since the $U(1)_R$ charge couples to mesonic operators and may lead to instabilities mentioned above, we will be interested in objects charged under the $U(1)_B$ symmetry only. The simplest such vertex operators are the $(N+1)^2$ products
${\cal A}_{m_1} {\cal \bar B}_{m_2}$ which are dual to combinations of D3-branes wrapping both the $(\theta_1,\phi_1,\psi)$ and $(\theta_2,\phi_2,\psi)$ directions. Turning on a chemical potential for $U(1)_B$ is expected to create a nonzero spatial density of such wrapped D3-branes. Our goal is to determine the background produced by them. We will use the simplifying assumption that the wrapped D3-branes are appropriately smeared over the $T^{1,1}$ coordinates, so that our solution will have the full $SU(2) \times SU(2)$ symmetry.

The $U(1)_B$ gauge field in $AdS_5\times T^{1,1}$ is contained in the components of the 4-form R-R-gauge field \cite{Klebanov:1999tb},
$C_4\sim A\wedge \omega_3$, where
\es{omegaDef}{
\omega_2 &\equiv {1\over 2} \left(\sin \theta_1 d\theta_1 \wedge d\phi_1
- \sin \theta_2 d\theta_2 \wedge d\phi_2 \right) \ , \\
\omega_3 &\equiv g_5 \wedge \omega_2 \ , \\
g_5 &\equiv d \psi + \cos \theta_1 d\phi_1 + \cos \theta_2 d\phi_2 \ .
}
Our ansatz for the self-dual 5-form field strength will therefore be\footnote{We define the Hodge dual of a $p$-form $\omega_p$ in $d$ space-time dimensions to be $(*\omega)_{i_1 i_2 \ldots i_{d-p}} = {1 \over p!} \epsilon^{j_1 j_2 \ldots j_p}_{\phantom{j_1 j_2 \ldots j_p}i_1 i_2 \ldots i_{d-p}} \omega_{j_1 j_2 \ldots j_p}$, where in a frame basis $\epsilon_{12\ldots d} = 1$.  This is equivalent to requiring $\omega \wedge *\omega = \abs{\omega}^2 {\rm vol}$, where $\abs{\omega}^2\equiv {1 \over p!} g^{i_1 j_1} \cdots g^{i_p j_p} \omega_{i_1 \ldots i_p} \omega_{j_1 \ldots j_p}$. \label{HodgeDef}}
\es{F5}{
F_5 &= {1 \over g_s} \left( {\cal F} + *{\cal F} \right) \ , \\
{\cal F} &= {2 L^4 \over 27} \omega_2 \wedge \omega_3 + {L^3 \over 9 \sqrt{2}} F \wedge \omega_3 \ ,\\
*{\cal F} &= {4 \over L} e^{-20 \chi/3}  \vol_M + {L^2 \over 3 \sqrt{2}} e^{2 \eta
- {4 \over 3} \chi} (*_M F) \wedge \omega_2 \ .
}
The normalization of the terms involving $F$ has been chosen so that the kinetic term for $F$ in the effective five-dimensional action is normalized canonically in the ultraviolet. Quantization of the 5-form flux requires that \cite{Herzog:2001xk}
\be
L^4 = 4 \pi g_s N (\alpha')^2 \frac{27}{16} \ .
\ee

At non-linear order, we expect the additional components of $F_5$ to cause a violation of the Poincar\'e invariance of the 5d metric, and also to produce a squashing of the internal space $T^{1,1}$.  A minimal consistent truncation of type IIB supergravity that contains these effects turns out to be
\es{10dMetric}{
ds_{10}^2 = e^{-{5\over 3} \chi} ds_M^2 + L^2 e^{\chi}
\bigg[{e^{\eta} \over 6}\sum_{i=1}^2 \left( d \theta_i^2 + \sin^2 \theta_i d\phi_i^2 \right)
+ {e^{-4 \eta} \over 9} g_5^2 \bigg] \ .
}
If $\chi = \eta = 0$, this metric reduces to the direct product between the non-compact space $M$ and $T^{1, 1}$.  The scalar $\chi$ controls the overall size of $T^{1, 1}$, while $\eta$ introduces a stretching of the $U(1)$ fiber relative to the two two-spheres.  These scalar fields in $AdS_5$ are dual to operators of conformal dimension $8$ and $6$, respectively.

The above ansatz yields a consistent truncation of type IIB SUGRA with the effective five-dimensional lagrangian
\es{Lag5d}{
{\cal L}_{\rm eff} &= R - {1 \over 4} e^{-{4 \over 3} \chi + 2 \eta} F_{\mu\nu}^2
- 5 (\partial_\mu \eta)^2 - {10 \over 3} (\partial_\mu \chi)^2 - V(\eta, \chi) \ ,
}
where the potential for the two neutral scalars is given by
\es{GotScalarPot}{
V(\eta, \chi) &\equiv  {8 \over L^2} e^{-{20\over 3} \chi}
+ {4 \over L^2} e^{-{8 \over 3} \chi} \left(e^{-6 \eta} - 6 e^{-\eta} \right) \ .
}
The scalar kinetic terms and potential had been previously determined in \cite{Klebanov:2000nc}.  However, the scalar coupling to the $U(1)_B$ gauge fields was not considered there. Indeed, \eqref{Lag5d} shows that the gauge kinetic term depends on the scalars; to study the baryonic black holes we need to include the squashing of $T^{1,1}$.

A seven-dimensional analogue of $T^{1,1}$ is the Sasaki-Einstein space $Q^{1,1,1}$ \cite{D'Auria:1983vy}.  The background $AdS_4\times Q^{1,1,1}$ has two kinds of baryonic symmetries corresponding to two topologically non-trivial 5-cycles that can be wrapped by M5-branes. In Appendix \ref{sec:Qbranes} we consider black 2-branes carrying one of these baryonic charges, and find a consistent truncation which includes a squashing of $Q^{1,1,1}$.

\section{Equations of Motion}
\label{sec:blackholes}

Using \eqref{Lag5d}, we will look for time-independent charged black 3-brane solutions.  We impose rotation and translation symmetry in the 3 spatial directions, but the Poincar\' e symmetry is obviously broken. Therefore we will use the ansatz
\es{5dansatz}{
ds_M^2 = -g e^{-w} dt^2 + \frac {dr^2} g + \frac {r^2} {L^2} \sum_{i=1}^3 (d x^i)^2 \ ,
}
where $g$ and $w$ are functions of $r$. This choice of parametrization (see, for example, \cite {Horowitz:2009ij}) will prove useful in simplifying the form of the equations of motion.  To turn on the baryonic charge density and chemical potential, we need to consider only the time component of the $U(1)_B$ gauge field, $A= \Phi(r) dt$, so that the field strength is  $F=dA= \Phi' dr\wedge dt$. We will also assume that the scalars $\chi$ and $\eta$ depend on the radial coordinate $r$ only.

We should note that our full ten-dimensional ansatz \eqref{F5}--\eqref{10dMetric} preserves the $\mathbb{Z}_2$ space-time inversion symmetry where $(t, \vec{x}) \rightarrow (-t, -\vec{x})$ is accompanied by the interchange of the two 2-spheres, $(\theta_1,\phi_1)\leftrightarrow (\theta_2,\phi_2)$. The forms $\omega_3$ and $\omega_2$ change sign under this transformation, but the terms $dt \wedge dr \wedge \omega_3$ and $dx^1\wedge dx^2\wedge dx^3\wedge \omega_2$ present in $F_5$ are invariant. In the gauge theory, this $\mathbb{Z}_2$ symmetry appears to correspond to $(t, \vec{x}) \rightarrow (-t, -\vec{x})$ accompanied by $A_i \leftrightarrow \bar B_i$.

With this ansatz, the effective one-dimensional lagrangian is
\es {Lag5dWithAnsatz} {
L_{\text{eff}} &= -5 r^3 g e^{-{1\over 2}w} \Bigl(\eta'{}^2 + \frac {2} 3 \chi'{}^2 \Bigr) +\frac 12 r^3 e^{{1\over 2}w} e^{2\eta - \frac 43\chi} \Phi'{}^2 - 3 (g r^2)' e^{-{1\over 2}w} - r^3 e^{-{1\over 2}w} V\ .
}
The equations of motion following from this lagrangian are
\begin{subequations} \label{5deoms}
\begin {align}
\chi'' + \chi' \left(\frac 3r + \frac {g'}g + \frac {5r} 3 \eta'^2 + \frac {10r} 9 \chi'^2 \right)
- \frac {\Phi'{}^2} {10 g} e^{w + 2 \eta - \frac 43 \chi}
- \frac 3{20 g} \frac {\partial V} {\partial \chi} &= 0 \ ,
\label {eomchi}\\
\eta'' + \eta' \left(\frac 3r + \frac {g'}g + \frac {5r} 3 \eta'^2 + \frac {10 r}9 \chi'^2 \right)
+ \frac {\Phi'{}^2} {10 g} e^{w + 2 \eta - \frac 43 \chi}
- \frac 1 {10 g} \frac {\partial V} {\partial \eta} &=0 \ ,
\label {eometa}\\
g' + g \left(\frac 2r + \frac {5r} 3 \eta'^2 + \frac {10 r}9 \chi'^2 \right)
+ \frac {r  \Phi'{}^2} {6} e^{w + 2 \eta - \frac 43 \chi}
+ \frac r3 V &=0 \ ,
\label {eomg}\\
\Phi'' + \Phi' \left(\frac 3r + \frac 12 w' + 2 \eta' - \frac 43 \chi' \right) &= 0 \label {eomphi} \ , \\
w' + \frac {10r}9 \bigl(3 \eta'^2 + 2 \chi'^2\bigr) &= 0 \label {eomw}\ ,
\end {align}
where primes denote derivatives with respect to $r$, and $V$ is the scalar potential defined in \eqref{GotScalarPot}.

Equation~\eqref{eomphi} can be integrated once yielding
\es{PhieomSimp}{
\Phi' &= \frac {Q} {r^3} e^{-\frac 12 w - 2 \eta + \frac 43 \chi} \ ,
}
\end{subequations}
where $Q$ is an integration constant related to the charge of the black hole. This is the conservation equation for baryonic charge. From now on, we will use \eqref{PhieomSimp} instead of \eqref{eomphi}. Plugging \eqref{PhieomSimp}
into equations~\eqref {eomchi}--\eqref {eomg}, we get the added bonus of eliminating the dependence on $w$, so the remaining equations are
\begin {subequations}
\label {eomsDMA}
\begin {align}
\chi'' + \chi' \left({3 \over r} + {g' \over g} + {5r \over 3} \eta'^2 + {10r \over 9} \chi'^2 \right)
- {Q^2 \over 10 r^6 g} e^{-2 \eta + {4 \over 3} \chi}
- {3 \over 20 g} {\partial V \over \partial \chi} &= 0 \ ,
\label {eomsDMAchi}\\
\eta'' + \eta' \left({3 \over r} + {g' \over g} + {5r \over 3} \eta'^2 + {10 r \over 9} \chi'^2 \right)
+ {Q^2 \over 10 r^6 g} e^{-2 \eta + {4 \over 3} \chi}
- {1 \over 10 g} {\partial V \over \partial \eta} &=0 \ ,
\label {eomsDMAeta}\\
g' + g \left({2 \over r} + {5r \over 3} \eta'^2 + {10 r \over 9} \chi'^2 \right)
+ {Q^2 \over 6 r^5} e^{-2 \eta + {4 \over 3} \chi}
+ {r \over 3} V &=0
\label {eomsDMAg}\ .
\end {align}
\end {subequations}
To reduce the system from five to three coupled differential equations was the main
motivation for using the ansatz \eqref{5dansatz}.

\section{Numerical Solutions}
\label{sec:numerics}

Solving the coupled non-linear equations \eqref{eomchi}--\eqref{PhieomSimp} is in general a difficult task. We will mostly rely on numerical work, but in some limits will be able to present analytical formulae. The simplest situation is when $Q=0$ and we find the well-known black 3-brane solution in $AdS_5$. In this solution, the shape of $T^{1,1}$ does not depend on $r$, i.e.\ the scalars $\chi$ and $\eta$ vanish. For small values of $Q$ we can use perturbation theory in this small parameter to obtain an analytic expansion of the solution. This exercise is carried out in Appendix~\ref{sec:smallcharge} where we find that the scalars $\chi$ and $\eta$ are now of order $Q^2$ and acquire a dependence on $r$.  In the next section, we present some analytical results in the extremal limit.  However, for intermediate values of $Q$, we know of no good analytical methods and resort to numerical ones.

\subsection {Setup for Numerics}

Finite-temperature solutions are found numerically by a standard shooting technique. The numerical solver is seeded close to the boundary, which is located at $r\to \infty$, by using a series expansion that imposes the correct boundary conditions.

The first task is to determine the boundary conditions. All our ten-dimensional metrics must asymptote to $AdS_5 \times T^{1, 1}$ at large $r$. This means that the asymptotic boundary conditions that we require as $r \to \infty$ are
\es{BdryConds}{
w &\to 0 \ , \qquad g = \frac {r^2} {L^2} + \mathcal {O} (L^2/r^2) \ ,\\
\eta &\to 0 \ , \qquad \chi \to 0 \ .
}
In order to describe states in the dual field theory at nonzero baryon chemical potential, we also require $\Phi \to \Phi_0$, for a constant $\Phi_0$ that will be related to the chemical potential shortly. Generically, solutions satisfying these boundary conditions will have an event horizon at some value $r = r_h$ where $g(r)$ vanishes.  The standard boundary conditions required at the horizon are that $w$, $\eta$, and $\chi$ should be finite.

From \eqref{eomchi}--\eqref{PhieomSimp} one can work out a series solution at large $r$ that satisfies the boundary conditions~\eqref{BdryConds}:
\es{BdySeries}{
w&= \mathcal {O} \left((L/r)^{12} \log^2 (r/L) \right) \ , \\
g &= {r^2 \over L^2} + \frac {g_2 L^2} {r^2} + \frac {Q^2 L^4} {12 r^4} + \mathcal {O} \left((L/r)^{10} \log^2 (r/L)\right) \ ,\\
\Phi &= \Phi_0 - \frac {Q} {2r^2} + \mathcal {O} \left((L/r)^{8} \log (r/L)\right) \ , \\
\chi &= -\frac {Q^2 L^2} {200 r^6} + \frac {\chi_8 L^8} {r^8} + \mathcal {O} \left((L/r)^{10}\right) \ , \\
\eta &= \frac {Q^2 L^2 \log (r/L)} {80 r^6} + \frac {\eta_6 L^6} {r^6} +\mathcal {O} \left((L/r)^{10} \log (r/L)\right)  \ .
}
All higher order terms in the series are determined in terms of $g_2$, $\Phi_0$, $\chi_8$, $\eta_6$, and $Q$.

To proceed further, it is useful to review some of the symmetries of our ansatz. The equations of motion \eqref{5deoms} and the boundary conditions are invariant under some scaling symmetries which act with the charges summarized in table~\ref{ScalingSymm}.  We say that a quantity $X$ has charge $q$ under a scaling symmetry if
\es{ChargeDef}{
X \to \lambda^q X \ .
}
\TABULAR {c||c|c|c|c|c|c|c|c|c}
{
Symmetry & $e^{-w}$ & $g$ & $\Phi$ & $\eta$ & $\chi$ & $t$ & $\vec{x}$ & $r$ & $L$ \\
\hline \hline
type A & $0$ & $0$ & $0$ & $0$ & $0$ & $1$ & $1$ & $1$ & $1$ \\
type B & $0$ & $2$ & $1$ & $0$ & $0$ &$-1$ &$-1$ & $1$ & $0$
}
{Charges under the scaling symmetries of the equations of motion \eqref{5deoms} and of the boundary conditions \eqref{BdryConds}.  These charges are defined as in \eqref{ChargeDef}.\label{ScalingSymm}}
The first symmetry in table~\ref {ScalingSymm} is a formal way of expressing the arbitrariness of a choice of units in the bulk. It can be used to set $L = 1$. The second symmetry can be used to put $r_h = 1$, but when shooting from the boundary it is more useful to employ the same symmetry to set $g_2 = -1$ instead.  We can also set $\Phi_0 = 0$, by using the fact that the potential $\Phi$ appears only through its derivatives in the equations of motion, and does not appear in the metric.  Having eliminated $\Phi_0$ and fixed $Q$, we are left with two parameters, $\chi_8$ and $\eta_6$, that need to be tuned in order to match to regular solutions at the horizon. Trial values for these parameters are determined from the small $Q$ (large $T$) expansion of Appendix A, and the results are used to go to progressively smaller temperatures.

Thermodynamic quantities such as the energy density $\epsilon$, entropy density $s$, temperature $T$, charge density $\rho$, and chemical potential $\mu$, can be computed from the following formulae:
\es{GotThermo}{
\epsilon &= -{3 \over 2} {g_2 \over \kappa_5^2 L} \ , \qquad
s = {2 \pi r_h^3 \over \kappa_5^2 L^3}  \ , \qquad
T = \frac {g'(r_h) e^{-w_h/2}} {4\pi} \ , \\
\rho &= {Q \over 2 \kappa_5^2 L^2} \ , \qquad \mu = {\Phi_0 - \Phi_h \over L} \ ,
}
where $\kappa_5$ is the five-dimensional gravitational constant, $w_h \equiv w(r_h)$, and $\Phi_h \equiv \Phi(r_h)$.
Using the parameters from \cite{Herzog:2001xk} we find
\es{5dk}{
{1\over \kappa_5^2}= {27 N^2\over 64 \pi^2 L^3}\ .
}
The energy density can also be computed from
\es{EpsilonVar}{
\epsilon = {3 \over 4} \left(T s + \mu \rho \right) \ ,
}
which follows from $\epsilon = T s - p + \mu \rho$ and the tracelessness of the stress tensor, $\epsilon = 3p$.

\subsection {Results}

Using the shooting method described above, we were able to find numerical black hole solutions for a fairly large range of temperatures.  Here are the main features we observe:
\EPSFIGURE {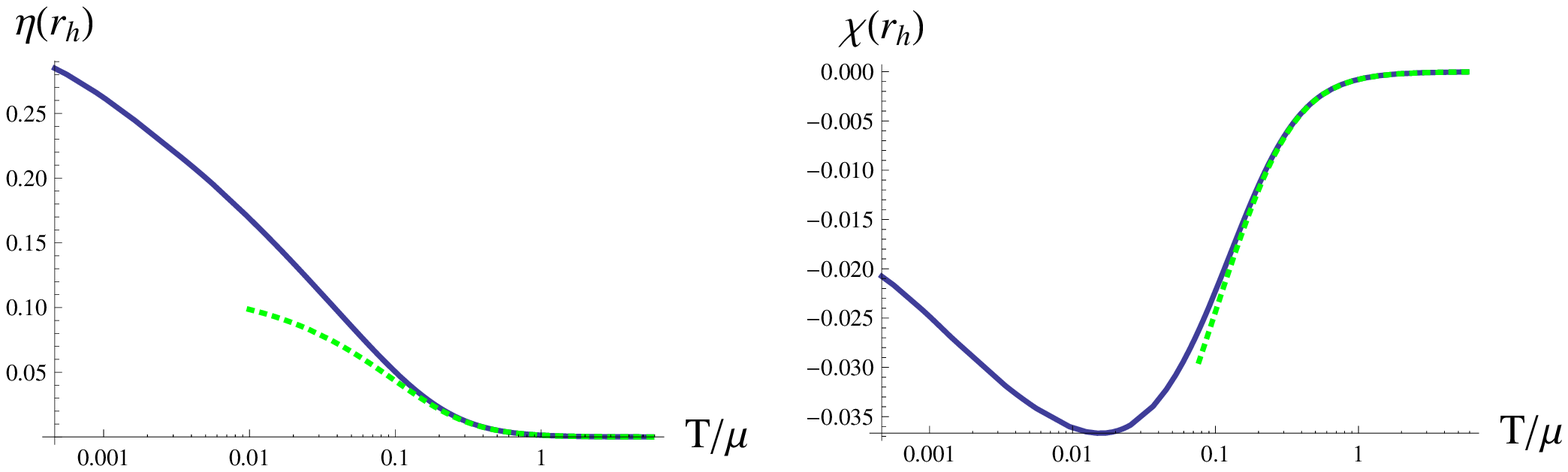, width=\textwidth} {The behavior of the horizon values of the scalars as a function of $T/\mu$. The dotted line is the high-temperature behavior obtained from the low $Q$ expansion in appendix~\ref{sec:smallcharge}.  According to the zero-temperature expansion developed in section~\ref{sec:zeroT}, $\eta_h$ and $\chi_h$ should diverge at $T = 0$.  We believe we are far from this regime.\label {fig:scalars}}
\begin{itemize}
\item Our numerical results agree with the analytical computations presented in Appendix~\ref {sec:smallcharge} in the limit of high temperatures.  See figure~\ref{fig:scalars} for a comparison of the values of the scalars at the horizon found numerically to those predicted by the analytic formula \eqref{etachiHor}.

\item We were able to construct numerical black hole solutions only for temperatures higher than $T \approx 0.0005 \mu$ because of loss of numerical precision at lower temperatures.  We believe the lowest temperatures we attained are not low enough to provide a thorough check of the analytical zero-temperature horizon expansion constructed in the next section.  In fact, we will argue towards the end of the next section that we expect the zero-temperature expansion to become a good approximation to the near-extremal solutions when $\log \log \mu/T \gg 1$, which is beyond the range of temperatures where our numerics are reliable.
\item
As one decreases the temperature, the Bekenstein-Hawking entropy seems to approach a non-zero value (see figure~\ref{fig:entropy}).  This fact will be confirmed analytically in the next section, where we find that as $T$ approaches zero, $s/\rho \approx 2.09$.  However, as will be discussed in the next section, there are stringy effects that make our solution trustworthy only down to an exponentially low temperature of order $\mu e^{-\text{const.} \times (g_s N)^{2/3} }$.
\EPSFIGURE {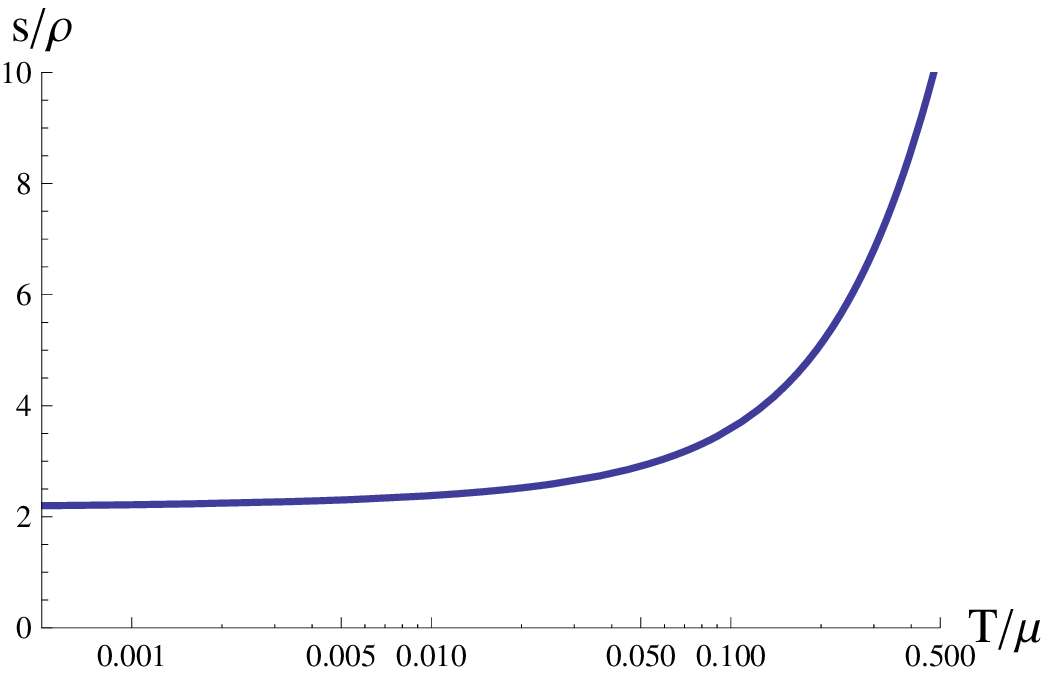,width=0.6\textwidth} {Entropy density normalized by charge density, as a function of temperature over chemical potential. The entropy is seen to go to a nonzero constant as $T\to 0$, in agreement with the zero-temperature expansion discussed in section~\ref{sec:zeroT}.\label {fig:entropy}}

\item Lastly, one may worry that at low enough temperatures the curvature of these backgrounds might get large and the supergravity approximation might break down.  As can be seen from figure~\ref{fig:curvature}, the horizon value of the ten-dimensional Riemann tensor squared is uniformly bounded from above.  In fact, in the following section we will show that all curvature invariants should vanish at the extremal horizon.
\EPSFIGURE {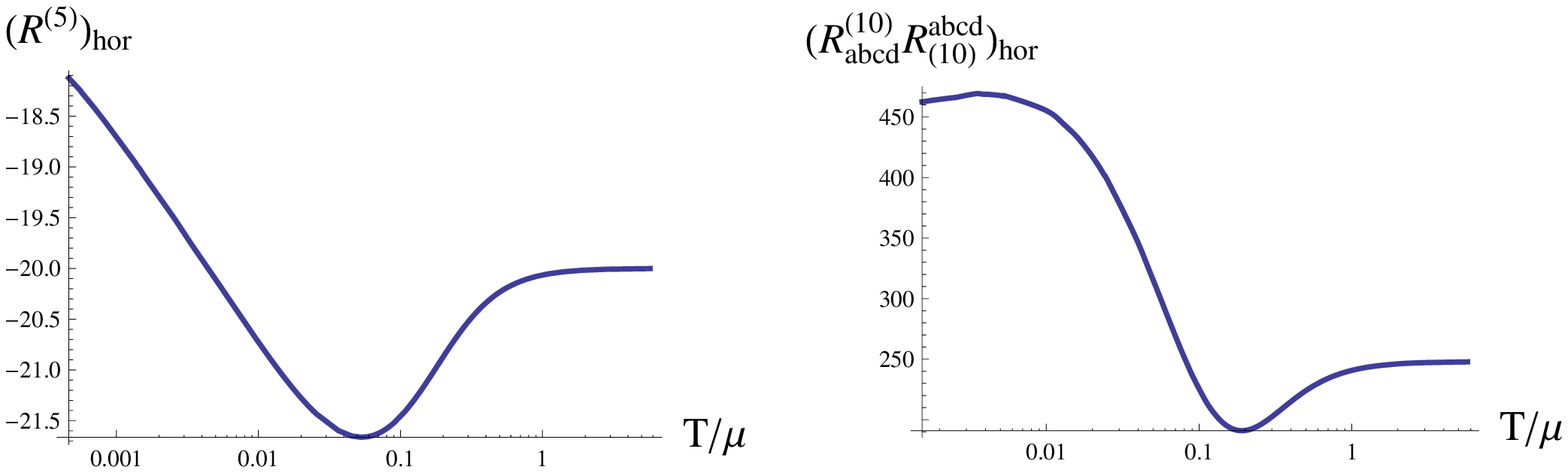,width=\textwidth} {The five-dimensional Ricci scalar (left), and the square of the ten-dimensional curvature (right) at the horizon, in units where $L = 1$, as a function of $T/\mu$.  As discussed in section~\ref{sec:zeroT}, all curvature invariants evaluated at the extremal horizon vanish.\label{fig:curvature}}
\end{itemize}

\section {Near-$AdS_2$ Near Horizon}
\label{sec:zeroT}

We now find a horizon series expansion at zero temperature.  We will restrict ourselves to the set of equations \eqref{eomsDMA}, since one can always use \eqref{eomw} and \eqref{PhieomSimp} to find $w(r)$ and $\Phi(r)$ afterwards.  In the following, we use the symmetries in Table~\ref {ScalingSymm} to set $L = r_h = 1$.

A guess for the zero-temperature value of the charge $Q$ that appears explicitly in equations \eqref{eomsDMA} can be found from the following line of reasoning based on properties of the nonzero temperature solutions.   At the horizon, $g(r_h = 1) = 0$; since $g$ must be positive outside the horizon, we need $g'(1) \ge 0$. Evaluating eq.~\eqref {eomg} at the horizon, we get
\es {gpAtHorizon} {g' (1) = -\frac 16 \,e^{-6 \eta_h - \frac {20}3 \chi_h}
\bigl(16 e^{6 \eta_h} + 8 e^{4 \chi_h} - 48 e^{5\eta_h + 4 \chi_h}
+ Q^2 e^{4\eta_h + 8\chi_h}\bigr) \ge 0\ ,
}
\newcommand {\expeta} {a}
\newcommand {\expfchi} {b}
which implies
\es{gpRoots}{
\expfchi_1 &\leq e^{4 \chi_h} \leq \expfchi_2 \ ,
}
where
\es {gpRoots_b} {
\expfchi_{1,2} = \frac 4 {Q^2 e^{4\eta_h}} \Bigl(-1 + 6 e^{5\eta_h} \pm
\sqrt {1 - 12 e^{5\eta_h} - (Q^2 - 36) e^{10\eta_h}}\Bigr)\ .
}
For $Q>6$, both $b_1$ and $b_2$ are negative or complex for any $\eta_h$, resulting in no possible range of $\chi_h$. For $Q<6$, the smallest positive value of $\eta_h$ for which $b_{1,2}$ are real is the one for which $b_1 = b_2$.   The fact that there are no solutions for $Q > 6$ implies from \eqref {GotThermo} that
\es {finiteEntropy} {
\frac s \rho = \frac {4\pi} {Q} \ge \frac {2\pi} 3 \approx 2.09\;,
}
an inequality that should hold at all $T$ within the supergravity approximation.

When $b_1 = b_2$, we have
\es {gpRootsCritical} {
\eta_c = -\frac{1}{5} \log (6 - Q) \ ,
\qquad \chi_c = -\frac{1}{20} \log(6 - Q) - \frac{1}{4} \log {Q \over 4}\ ,
}
which implies that $\eta_c$ and $\chi_c$ go to infinity as $Q \to 6$, while
\es {abFiniteLimit} {
\eta_c - 4 \chi_c = \log \frac {Q} 4 \to \log \frac{3}{2} \quad \text{ as $Q \to 6$.}
}
A reasonable guess is that at zero temperature $Q = 6$ and the scalars $\chi$ and $\eta$ diverge at the extremal horizon, while $\eta - 4 \chi$ approaches $\log{3 \over 2}$.

With $Q = 6$, we use the method of dominant balance to find the asymptotic behavior of zero-temperature solutions at the horizon.  By adding arbitrary power series to these dominant terms, we find we can satisfy the equations of motion. We obtain a solution of the form
\newcommand {\newr} {\tilde r}
\es {GotLeading} {
\chi &= -\frac 1 {20} \log \biggl(\frac {2187} {16} \newr\biggr) + \frac {1063} {1000} \newr + \dotsb \ , \\
\eta &= -\frac 1 5 \log (18 \newr) + \frac {463} {250} \newr + \dotsb \ , \\
g &= \newr^{13/3} \biggl(\frac {93312 \cdot (12)^{1/3}} {25} + \dotsb \biggr) \ , \\
w & = \frac 5 {36 \newr} + \frac {77} {30}\log \newr + \tilde w + \dotsb \ ,
}
where
\es {Defs} {
\newr &\equiv r - 1\ ,
}
and the dots stand for regular Taylor series.

As $\newr \to 0$, $g_{00}=-g \,e^{-w}$ has an essential singularity. This prompts us to introduce a better coordinate,
\es {newCoord} {
y= 3^{-{193/ 120}} 2^{-{107/ 120}} 5^{3 / 40} e^{\tilde w/2} \,e^{5 / (72 \newr)} \ ,
}
which becomes large near the horizon. In terms of this coordinate, and reinstating the factors of $L^2$, the near-horizon metric becomes
\es{10dMetric_b}{
ds_{10}^2 &=
\frac {L^2} {y^2} \Bigl(- (\log y)^{-37/20} dt^2 + \frac 1 {6\,(30)^{1/4}} (\log y)^{1/4} dy^2\Bigr) + L^2\left(\frac {1215} {128}\right)^{1/12} (\log y)^{-1/12} d\vec x^2 \\
&\qquad + \biggl[\frac {L^2} {6} \biggl(\frac {15}8\biggr)^{-1/4} (\log y)^{1/4} \sum_{i=1}^2 \left( d \theta_i^2 + \sin^2 \theta_i d\phi_i^2 \right) + \frac {L^2} {9} \biggl (\frac {125} {96}\biggr) ^{1/4} (\log y)^{-3/4} g_5^2  \biggr] \ .
}
This is a warped product $AdS_2 \times {\mathbb R}^3\times T^{1,1}$ with warp factors that are powers of the logarithm of the AdS radius $y$.  The appearance of the logarithmic warp factors makes it a novel type of ``nearly conformal'' IR behavior.  Note that the 3d spatial components of the metric tend to zero in the IR as $g_{ij} \sim (\log y)^{-1/12} \delta_{ij}$; this is an important difference from the RNAdS metric where they approach a constant.  Remarkably, the extra logarithms actually reduce the curvature, so that we can trust the supergravity approximation everywhere: we have checked that the 10d Kretschmann invariant and the 5d Ricci scalar scale as $R_{abcd} R^{abcd} \sim (\log y)^{-1/2}$ and $R \sim (\log y)^{-1/3}$ respectively. Such an asymptotic reduction of curvature due to the appearance of logarithms also occurs in the UV region of the warped deformed conifold \cite{Klebanov:2000nc,Klebanov:2000hb}.

One can use the asymptotic solution to estimate what happens at temperatures so low that the horizon is located deep inside the near-$AdS_2$ region, at $\tilde r_h \ll 1$.  If we assume that \eqref{GotLeading} approximately holds down to the horizon,  we can estimate using \eqref{GotThermo} that $T/\mu \sim e^{-{5 \over 72 \tilde r_h}}$ up to power law corrections in $\tilde r_h$ and numerical factors.   From \eqref{GotLeading}, one can then compute the values of the scalars at the near-extremal horizon:
\es{ScalarsHorNonzeroT}{
\eta_h = 4 \chi_h &= -{1 \over 5} \log \tilde r_h + {\cal O}(\tilde r_h^0) =  {1 \over 5} \log \log {\mu \over T}  + {\cal O} (({T/ \mu})^{0}) \ .
}
We therefore expect the approach to the extremal limit to be very slow and hard to investigate numerically because one has to go to exponentially small temperatures.

This numerical suggestion of the existence of an exponentially small scale compared to the chemical potential $\mu$ in the conifold gauge theory is corroborated by a stringy argument.
In the type IIB string theory context, our solution cannot be trusted for arbitrarily low temperatures.\footnote{We are very grateful to J. Maldacena and E. Witten for pointing this out to us.} In the $T=0$ solution the $\psi$-circle shrinks to zero size at the horizon (see the last term of \eqref{10dMetric_b}), and the standard approach is to T-dualize along this direction for $\log y > \text{const.} \times (g_s N)^{2/3}$ where the size of the circle becomes of the order of the string length $\sqrt{\alpha'}$.
Estimating the temperature at which the size of the $\psi$-circle at the horizon reaches the string scale, we find
\es{limtemp}{\log  {\mu \over T} \sim (g_s N)^{2/3}
\ .
}

It is remarkable that application of string theoretic arguments to our black brane suggests that in the conifold gauge theory at baryon chemical potential $\mu$ there exists an energy scale of order $\mu e^{-\text{const.} \times (g_s N)^{2/3} }$.
From the point of view of the type IIB $\sigma$-model with coupling $\sim (g_s N)^{-1/4}$,
such a scale can arise only non-perturbatively.  Below this exponentially small scale, the gauge theory presumably exhibits some new effects that can be studied by T-dualizing the solution and lifting it to M-theory. While the background \eqref{GotLeading} has a non-vanishing Bekenstein-Hawking entropy and is smooth, stringy effects become important when the $\psi$-circle becomes small, and these effects could remove the conflict with the conventional statement of the Third Law of Thermodynamics \cite{Heemskerk:2009pn}. Since the type IIB backgrounds can be trusted when the $\psi$-circle is above the string scale, we expect the entropy to be approximately constant for 
$\mu e^{-\text{const.} \times (g_s N)^{2/3} } < T\ll \mu$, i.e. when the horizon is inside the near-$AdS_2$ throat. It would be very interesting to provide a microscopic origin of this entropy of order $N^2$ by studying the underlying D3-brane system, which involves two intersecting stacks of D3-branes wrapped over the $T^{1,1}$. We hope to return to these issues in the future.

In order to match the asymptotic solution \eqref{GotLeading} to the appropriate behavior at the boundary, we generically need a two-parameter family of solutions, in addition to the integration constant $\tilde w$, which is related to a choice of units for time in the boundary theory. Following~\cite {Horowitz:2009ij}, we linearize around this leading behavior, to find non-analytic pieces that we might have missed. There are two solutions to the linearized equations, of the form
\es {GotLinearizedSolns} {
\delta \chi &= e^{\alpha/\newr} \newr^{c_1} (1 + \dotsb) \ , \\
\delta \eta &= 4 \delta \chi + e^{\alpha/\newr} \newr^{c_2} (1 + \dotsb) \ , \\
\delta g &= e^{\alpha/\newr} \newr^{c_3} (1 + \dotsb)\ ,
}
where the corrections are just series in positive powers of $\newr$. The values of $\alpha$ and $c_i$ for the two solutions are
\es {explicitSoln1} {
\alpha = -\frac 5{72} \ , \qquad c_1 = \frac 1{20} \ , \qquad c_2 = c_1 + 1 \ , \qquad c_3 = c_1 + \frac {10}3\ ,
}
and
\es {explicitSoln2} {
\alpha = -\frac 5 {144} \bigl (\sqrt {21} - 1\bigr) \ , \qquad c_1 = \frac {179 \sqrt {21} - 189} {360} \ , \qquad c_2 = c_1 \ , \qquad c_3 = c_1 + \frac {13}3 \ .
}

For a scalar in $AdS_2$, the behavior of linearized modes as a function of the radius $y$ is
\es{sourcevev}{
\phi (y,t)\sim \phi_0 (t) y^{1-\Delta_{\mathrm{IR}}} + A(t) y^{\Delta_{\mathrm{IR}}} \ ,
}
where $\phi_0(t)$ is the source for an operator of dimension $\Delta_{\mathrm{IR}}$, and $A(t)$ is its expectation value \cite{Klebanov:1999tb}. In the case at hand, since the background is $AdS_2$ up to powers of $\log y$, we expect
\eqref{sourcevev} to hold up to similar powers. The perturbations \eqref{GotLinearizedSolns} are indeed of this form.
The first of these solutions corresponds to a source for an operator of $\Delta_{\mathrm{IR}}=2$; the second to a source for
$\Delta_{\mathrm{IR}}=(1+\sqrt{21})/2$. Both of these operators are irrelevant in the nearly conformal quantum mechanics found in the IR; hence, they produce perturbations that decay at large $y$.

Another type of perturbation that we can study easily is a minimally coupled massless scalar, for example the dilaton.  In the near-$AdS_2$ region, the minimal scalar solutions behave as $a + b\, y (\log y)^{4/5}+ \dotsb$. We identify the dual operator as a marginal operator with $\Delta_{\mathrm{IR}}=1$. In the case of the dilaton, the operator is exactly marginal because the string coupling is a parameter in our solution.

The extremal solution described above is reminiscent of the ``run-away'' attractor flows described in \cite{Goldstein:2009cv}.  Typically, an attractor flow is a set of solutions to the supergravity equations of motion where the ultraviolet behavior is not universal, but in the infrared the scalars approach fixed values.  In the run-away case at least one of these fixed values is at infinity.  Our extremal solution would be of run-away type because the scalars $\eta$ and $\chi$ diverge at the extremal horizon, even though the combination $\eta - 4 \chi$ stays finite.  An important difference between the extremal solutions from \cite{Goldstein:2009cv} and the one we found is that in \cite{Goldstein:2009cv} the entropy density vanishes at extremality while in our case it does not.

\section{Baryonic Operators and D3-brane Probes}
\label{sec:noPT}

In this section
we investigate whether the baryonic black branes constructed above are stable with respect to condensation of baryonic operators in the conifold gauge theory.  As was already described, these operators are dual to wrapped D3-branes in the dual supergravity.
These wrapped D3-branes act like charged particles; there is a competition between the gravitational attraction and electrostatic repulsion between the particle and the charged black brane.
To test stability, we propose a simple thought experiment.  If we place such a wrapped D3-brane into the geometry and the brane falls into the black hole, we conclude the black hole is stable. However, if the brane finds some meta-stable minimum outside the horizon, we conclude the black hole is unstable and more such wrapped D3-branes can bubble off the horizon, find their way to the minimum in the potential, and reduce the baryonic charge on the black hole.  It is possible that more exotic bound states of D-branes may lead to instabilities.  We leave an investigation of such issues to future work.

\subsection{Charged Particles in Reissner-Nordstr\"om AdS}
\label{subsec:particles}

Instead of beginning with the D3-brane case, let us first consider the simpler case of a
particle of mass $m$ and charge $q$.
The action for such a particle is
\es{ParticleAction}{
S_p = -m \int ds\, \left\lvert{dx^\mu \over ds}\right\rvert - q \int ds\, A_\mu {d x^\mu \over ds} \ .
}
We restrict to a metric of the form (\ref{5dansatz}) with a radial electric potential $\Phi(r) = A_t(r)$.  For a particle sitting at some fixed $r$ and $\vec x$, the action becomes
\es{PActionRNAdS}{
S_p = - \int dt\, \left[m e^{-w(r)/2} \sqrt{g(r)} + q \Phi(r)  \right]
}
in the gauge $s = t$.  Equation \eqref{PActionRNAdS} shows that the potential for this particle as a function of $r$ is given by
\es{Potential}{
V(r) = m e^{-w(r)/2} \sqrt{g(r)} + q \Phi(r) \ .
}
The first term is the gravitational attraction of the black hole, and the second term the electrostatic repulsion.

To proceed further, we need explicit results for $w$, $g$, and $\Phi$.  We will return to the baryonic black hole momentarily, but let us first consider what happens for the simpler case of a Reissner-Nordstr\"om black hole with negative cosmological constant.  Given the Einstein-Maxwell action
\be
S_{EM} = \frac{1}{2 \kappa^2} \int d^{d+1} x \sqrt{-g} \left[ R + \frac{d(d-1)}{L^2} \right]
- \frac{1}{4 e^2} \int d^{d+1} x \sqrt{-g} F_{\mu\nu} F^{\mu\nu} \ ,
\ee
there is a black-brane solution with a horizon at $r = r_h$ and with $w=0$,
\be \label{gRNAdS}
g = \frac{r^2}{L^2} \left( 1 + Q^2 \left( \frac{r_h}{r} \right)^{2d-2} - (1+Q^2) \left( \frac{r_h}{r} \right)^d \right) \ ,
\ee
\be
\Phi(r) = Q \frac{e}{\kappa} \frac{r_h}{L} \sqrt{ \frac{d-1}{d-2}} \left[ \left(\frac{r_h}{r}\right)^{d-2} - 1 \right] \ .
\ee
The corresponding Hawking temperature is
\be
T_H = \frac{d-(d-2)Q^2}{4\pi} \frac{r_h}{L^2} \ .
\ee
Without loss of generality, let's assume $Q>0$.  The RNAdS black hole becomes extremal at
$Q=Q_{\rm max} \equiv \sqrt{\frac{d}{d-2}}$ where $T_H$ vanishes.

Using these formulae for $g$ and $\Phi$, one can check that for values of $q/m < q_{\rm crit}/m$, $V(r)$ is increasing monotonically for all values of $Q$ smaller than $Q_{\rm max}$, while for large values $q/m>q_{\rm crit}/m$, $V(r)$ has a minimum at some $r = r_*(Q)$ provided that the charge $Q$ of the black hole is larger than some critical value $Q_c$ that depends on $q/m$.
An expression for the critical charge $q_{\rm crit}$ can be found by requiring that the minimum of $V(r)$ occurs right at $r = r_h$ when $Q = Q_{\rm max}$.   The equation $V'(r_h) = 0$ is easily solvable in this case, yielding
\es{Gotqcrit}{
{q_{\rm crit}} = m \frac{\kappa}{e} \ ,
}
independent of the number of dimensions of the RNAdS space.\footnote{
This ratio has some broader significance, as emphasized by \cite{ArkaniHamed:2006dz} in the context of a weak gravity bound.  If extremal
Reissner-Nordstr\"om black holes are to be able to decay, there must always exist at least one particle whose charge-to-mass ratio is greater than $\kappa/e$.
}

This picture of an instability is essentially the classical limit of the superfluid or superconducting instability studied in refs.\ \cite{Gubser:2008px,Hartnoll:2008vx}.  In these papers, the charged particle is replaced with a charged scalar field $\psi$.  One solution to the equations of motion is $\psi=0$, but for $Q>Q_c$, there exists a second, more stable branch of solutions with $\psi \neq 0$.  The classical analog of this second branch is a cloud of charged particles sitting at the minimum in $V(r)$ described above.  Moreover, for large $m$ and $q$, the classical and field theory results for $Q_c$ agree.

\subsection{The Wrapped D3-brane}

The probe D3-brane action takes the form
\begin{equation}
S_{D3} = - \mu_3 \int d^4x \, e^{-\phi} \sqrt{-\det G_{ab}} + \epsilon \, \mu_3 \int C_4 \ ,
\end{equation}
where $\mu_3 = (2 \pi)^{-3} (\alpha')^{-2}$, $\alpha'$ is related to the string tension, and $\phi$ is the dilaton.  For our supergravity solution, the dilaton is a constant we take to be related to the string coupling, $g_s = e^\phi$.  The parameter $\epsilon$ is equal to one, but we leave it arbitrary so that we can tune the charge of the D3-brane.  We assume the metric ansatz (\ref{10dMetric}) and (\ref{5dansatz}).

We assume the probe brane sits at constant $\theta_2$, $\phi_2$, $x_i$, and $r$ and wraps the remaining four directions, including time. From $S_{D3}$ and the ansatz for $F_5$ (\ref{F5}) we deduce a potential for the D3-brane:
\es{GotVT11}{
V(r) = \frac{3N}{4L}  e^{-\frac 12 w(r)-\eta(r) + \frac 23\chi(r)} \sqrt{g(r)} - \epsilon \frac{3N\sqrt{2}}{8 L} \Phi(r) \ .
}

Comparing \eqref{GotVT11} with \eqref{Potential}, we see that in the UV where the scalars $\eta$ and $\chi$ are negligible, $e^{w(r)} \approx 1$ and $g \approx r^2/L^2$, the wrapped D3-brane corresponds to a particle in RNAdS with $m/q = \sqrt{2}$.  As shown at the end of section~\ref{subsec:particles}, this ratio corresponds to the critical value $m/q_{\rm crit}$ at which in RNAdS the potential $V(r)$ has a local minimum at $r = r_h$ in the extremal case.  There is no reason to expect that the critical value of $m/q_{\rm crit}$ should be the same in the presence of scalars, but we will now show that this is nevertheless true.

The simplest demonstration is to plot $V(r)$ numerically.  At the lowest temperatures we can access, there is no minimum in the potential when $\epsilon = 1$.  However, for $\epsilon > 1$ and $T$ sufficiently low, there is such a minimum, suggesting the D3-brane does indeed have this critical ratio of charge to mass.
Indeed, the minimum is observed to occur for $\epsilon>\epsilon_0$, where $\epsilon_0$ is some critical value larger than $1$. In figure~\ref {fig:condensation}, we plotted the dependence of this critical value on $T/\mu$. We see that as the temperature goes to zero, the critical value goes to $1$. The position of the minimum moves towards the horizon, showing that wrapped D3-branes have the marginal ratio of charge to mass that means they barely escape condensation.
\EPSFIGURE {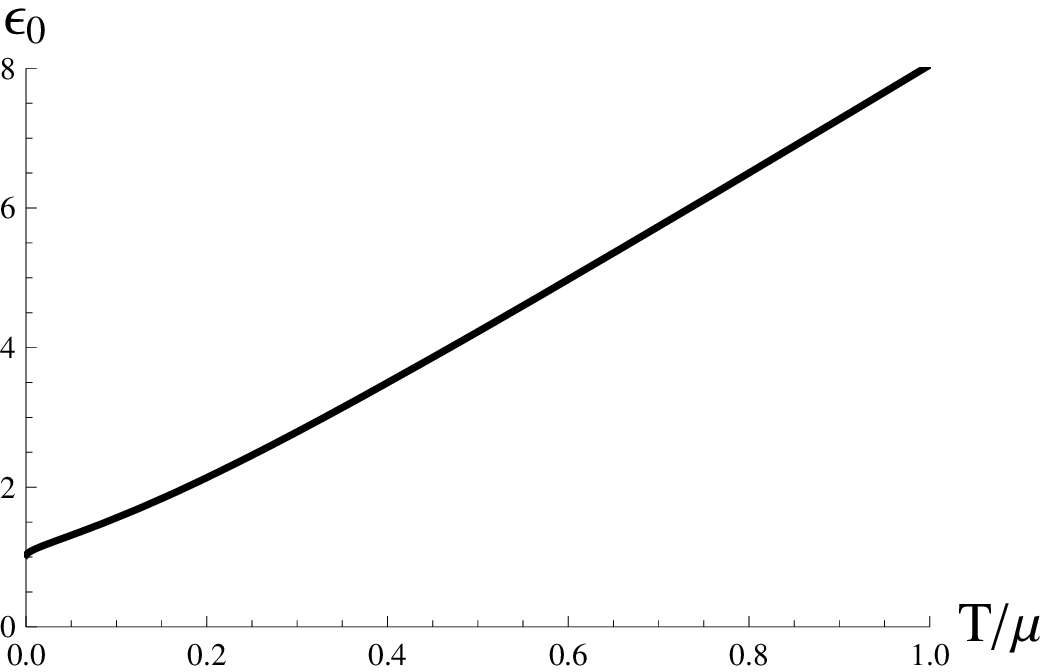, width=0.6\textwidth} {Critical value of the parameter $\epsilon$ above which condensation occurs, as a function of $T/\mu$. Just above $\epsilon_0$, the position of the potential minimum goes towards the horizon as $T \to 0$.\label {fig:condensation}}

Some analytic support for this claim comes from studying the conditions that the minimum in the potential disappear, $V'(r_*) = V''(r_*)=0$. Restoring $\epsilon =1$, these two vanishing conditions imply the relation
\be
e^{\eta(r_*)} = \frac{3}{2} e^{4 \chi(r_*)} \ .
\ee
However, we know this condition is satisfied in the $T\to 0$ limit at the horizon from (\ref{abFiniteLimit}).  Thus the
system just begins to become unstable at $T=0$, and we expect no phase transition.

More explicitly, we examine the potential (\ref{GotVT11}) in the $T\to 0$ limit.
From the zero-temperature series expansion (\ref{GotLeading}), we find that
\be
V(r) = N e^{-{5 \over 72 (r-1)} - {1 \over 2} \tilde w} (r-1)^{1/20} \left(
(1-\epsilon) \frac{324 \cdot 18^{1/6}}{5} (r-1) + O(r-1)^2 \right) \ ,
\ee
where we set $L = r_h = 1$.  Thus, the leading term in this series expansion vanishes when $\epsilon=1$.

\section{Another Stability Check}
\label{sec:stability}

The results of the previous section show that the black holes with baryonic charge we have constructed are stable against the simplest condensing operators that carry nonzero baryonic charge. Recall that all such operators have large dimensions of order $N$.  One may worry about a different kind of instability where at low enough temperatures there is a phase transition driven by operators that are uncharged under the baryonic symmetry.  That such a phase transition is in principle possible was noted in \cite{Hartnoll:2008kx,Denef:2009tp,Gubser:2009qm} for the case of the RNAdS black hole.  In that case, all uncharged operators with UV conformal dimension smaller than $3$ (when the gauge theory is $3+1$-dimensional) could trigger such an instability.

In this section we will study the stability of certain modes in the IR near-$AdS_2$ region of the $T=0$ solution. For a minimally coupled scalar,
\es{opdim}{\Delta_{\mathrm{IR}\pm} = {1\over 2} \pm \sqrt{{1\over 4} + m^2 L_{AdS_2}^2}
\ .
}
Hence, the Breitenlohner-Freedman (BF) stability bound in $AdS_2$ is
\es{BFViolation}{
m^2 L_{AdS_2}^2 \geq -{1\over 4} \ ,
}
and the dimension $\Delta_{\mathrm{IR}-}$ is allowed only in a narrow range of $m^2$ above this bound \cite{Klebanov:1999tb}. When the BF bound is violated the dimensions become complex, and the modes exhibit the oscillatory behavior as a function of the radius that is characteristic of ``bad tachyons'' (see for example \cite{Gibbons:2002pq}).
This simple analysis is not directly applicable to the case
of interest to us because the background is not exactly $AdS_2$, and the scalars are not minimally coupled. However, we will adopt a similar stability criterion. In the linearized approximation, an instability will be associated with the presence of complex dimensions and the ensuing oscillatory behavior of modes.

The baryonic black 3-branes studied in this paper come from a ten-dimensional type IIB construction, so one can study fluctuations of various (uncharged) supergravity fields and check their stability. A full analysis of all possible modes is beyond the scope of this paper.  However, we will demonstrate stability with respect to perturbations associated with a field theory operator of UV conformal dimension $\Delta = 2$. This operator is ${\rm Tr} (A_i \bar A_i - B_j \bar B_j)$, and its dual supergravity field is the ``resolution mode'' of the conifold, $\lambda$, that allows the two $S^2$'s to have different sizes. This is the most relevant mode that is odd under the $\mathbb{Z}_2$ space-time inversion symmetry accompanied by the interchange of the two 2-spheres. We thought that this mode was the most likely to cause an instability because it saturates the BF stability bound in $AdS_5\times T^{1,1}$; luckily, as we show, it does not destroy the stability of the near-$AdS_2$ solution.

The challenge here is that $\lambda$ mixes with the time components of certain gauge fields, already at the linearized level.  We were nevertheless able to find a consistent set of supergravity fields that include $\lambda$ and that decouple from all other fluctuations, providing a more general (non-linear) consistent truncation of type IIB supergravity than the one considered in section~\ref{sec:setup}.  Indeed, the consistent truncation we find reduces to the one in section~\ref{sec:setup} in a particular limit.  The full ten-dimensional ansatz and the effective five-dimensional action are given in section~\ref{sec:generaltruncation}.  In order to examine the stability of the near-$AdS_2$ geometry, in section~\ref{sec:fluctuations} we linearize the equations of motion around the baryonic black brane background and develop a horizon series expansion at zero temperature using the explicit solution discussed in section~\ref{sec:zeroT}.  Remarkably, the linearized equations contain mixings between the modes and effective mass terms that stabilize all potentially unstable modes within this ansatz.

\subsection{A More General Consistent Truncation}
\label{sec:generaltruncation}

A consistent truncation that extends \eqref{F5}--\eqref{10dMetric} to include the resolution mode of $T^{1, 1}$ can be constructed as follows. Compared to \eqref{F5}--\eqref{10dMetric}, this truncation has three additional fields:  the scalar field $\lambda$, which is the resolution mode of the conifold, and the spin-one fields $A_R$ and $\tilde A_R$, which mix to give a gauge field corresponding to the R-symmetry of the gauge theory as well as a massive spin-one field.  The metric ansatz is
\es{10dMetricPert}{
ds_{10}^2 &= e^{-5 \chi/3} ds_M^2 + L^2 e^{\chi}
\bigg[{e^{\eta + \lambda} \over 6} (d \theta_1^2 + \sin^2 \theta_1 d\phi_1^2) \\
{}&+ {e^{\eta-\lambda} \over 6} (d \theta_2^2 + \sin^2 \theta_2 d\phi_2^2)
+ {e^{-4 \eta} \over 9} \left(g_5 + {3 \over \sqrt{2} L} A_R\right)^2 \bigg] \ .
}
Defining
\es{FormDefs}{
g_5^A \equiv g_5 + {3 \over \sqrt{2} L} A_R \ ,
}
the self-dual five-form can be written as
\es{F5Pert}{
F_5 &= {1 \over g_s} \left( {\cal F} + *{\cal F} \right) \ , \\
{\cal F} &= {2 L^4 \over 27} \omega_2 \wedge \omega_2 \wedge g_5
+ {L^3 \over 9 \sqrt{2}} F \wedge \omega_2 \wedge g_5^A  \\
{}&- {L^3 \over 18 \sqrt{2}} \tilde F_R \wedge d g_5 \wedge g_5^A
+ {L^3 \over 18 \sqrt{2}} \tilde A_R \wedge dg_5 \wedge dg_5  \ , \\
*{\cal F} &= {4 \over L} e^{-{20 \over 3}\chi} \vol_M
+ {L^2 \over 3 \sqrt{2}} e^{-{4 \over 3} \chi + 2 \eta}
\left[\cosh(2 \lambda) *_M F - \sinh (2 \lambda) *_M \tilde F_R \right] \wedge \omega_2 \\
{}&+ {L^2 \over 6 \sqrt{2}} e^{-{4 \over 3} \chi + 2 \eta}
\left[\cosh(2 \lambda) *_M \tilde F_R - \sinh(2 \lambda) *_M F \right] \wedge dg_5 \\
{}&+ {2 \sqrt{2} \over 3} e^{-4 \chi - 4 \eta} *_M (A_R + \tilde A_R) \wedge g_5^A \ ,
}
where we have defined the field strengths $F = d A$, $F_R = d A_R$, and $\tilde F_R = d \tilde A_R$.

The effective five-dimensional action for this consistent truncation can be written as a sum of a bulk piece and a Chern-Simons term:
\es{EffectiveAction}{
S = \int d^5 x\, \sqrt{-g} {\cal L} + S_{\rm CS} \ .
}
The bulk lagrangian ${\cal L}$ is given by
\es{EffectiveLag}{
{\cal L} &= R - {10 \over 3} (\partial_\mu \chi)^2 - 5 (\partial_\mu \eta)^2
- (\partial_\mu \lambda)^2 - V(\eta, \chi, \lambda) \\
{}&- {1 \over 4} e^{2 \eta -  {4 \over 3} \chi} \bigg[
\cosh (2 \lambda) \left( F_{\mu\nu} F^{\mu\nu}
+ \tilde F^R_{\mu\nu} \tilde F_R^{\mu\nu} \right)
- 2 \sinh (2 \lambda) F_{\mu\nu} \tilde F_R^{\mu\nu}
\bigg] \\
{}&-{1\over 8} e^{-4 \eta + {8 \over 3} \chi} F^R_{\mu\nu} F_R^{\mu\nu}
- {4 \over L^2} e^{-4 \eta - 4 \chi} (A^R_{\mu} + \tilde A^R_{\mu})^2  \ ,
}
where
\es{GotV}{
V(\eta, \chi, \lambda) = {8 \over L^2} e^{-{20 \over 3} \chi} +
{4 \over L^2} e^{-{8 \over 3} \chi} \left( e^{-6 \eta} \cosh(2 \lambda) - 6 e^{-\eta} \cosh \lambda \right) \ .
}
The Chern-Simons part of the action is
\es{GotChernSimons}{
S_{\rm CS} = {1 \over 2 \sqrt{2}} \int \tilde A_R \wedge \tilde F_R \wedge F_R
- {1 \over 2 \sqrt{2}} \int A \wedge F \wedge F_R \ .
}

From now on, we restrict to a time-independent background where we have rotation and translation symmetry in the three non-compact directions $x^i$.  The most general ansatz with these symmetries is where the scalars $\chi$, $\eta$, and $\lambda$ depend only on $r$, and where
\es{RadialDependence}{
ds_M^2 &= -g e^{-w} dt^2 + \frac {dr^2} g + \frac {r^2} {L^2} \sum_{i=1}^3 (d x^i)^2 \ , \\
A &= \Phi(r) dt \ , \qquad A_R = \Phi_R(r) dt \ , \qquad
\tilde A_R = \tilde \Phi_R(r) dt \ ,
}
generalizing the ansatz used in section~\ref{sec:blackholes}.  In this case, the equations of motion following from the effective action \eqref{EffectiveAction} admit two conserved charges $Q_B$ and $Q_R$ associated to the baryonic symmetry and to the R-symmetry of the gauge theory, respectively.  The conservation equations take the form
\begin{subequations} \label{ConservedCharges}
\begin{align}
\Phi' \cosh (2\lambda) - \tilde \Phi_R' \sinh (2\lambda)
&= {Q_B \over r^3} e^{-{1 \over 2} w - 2 \eta + {4 \over 3} \chi} \ , \label{ConservedBCharge}\\
\tilde \Phi_R' \cosh (2\lambda) - \Phi' \sinh (2\lambda) - {1\over 2} e^{-6\eta + 4\chi} \ \Phi_R'
&= {Q_R \over r^3} e^{-{1 \over 2} w - 2 \eta + {4 \over 3} \chi} \ ,  \label{ConservedRCharge}
\end{align}
\end{subequations}
where, as usual, primes denote derivatives with respect to $r$.  The first of these two equations is a generalization of \eqref{PhieomSimp}. In the UV, the scalars are negligible and the above two equations reduce to $\Phi' \approx {1\over r^3} Q_B$ and $\tilde \Phi_R' - {1 \over 2} \Phi_R'  \approx {1 \over r^3} Q_R$, justifying the interpretation of $Q_B$ as the baryonic charge and of $Q_R$ as the R-charge.

\subsection{Horizon Expansion of Linearized Fluctuations}
\label{sec:fluctuations}

As mentioned in section~\ref{sec:blackholes}, our background is invariant under a $\mathbb{Z}_2$ symmetry that acts by flipping the sign of the non-compact coordinates, $(t, \vec{x}) \to (-t, -\vec{x})$, and interchanging the two spheres in the compact space, $(\theta_1, \phi_1) \leftrightarrow (\theta_2, \phi_2)$.  Recall that in the background solution only $\chi$, $\eta$, and $\Phi$ are nonzero.  The fluctuations around this background are distinguished by their parity properties under the $\mathbb{Z}_2$ symmetry:  $\delta \Phi$, $\delta \chi$, and $\delta \eta$ are even, while $\delta \Phi_R$, $\delta \tilde \Phi_R$, and $\delta \lambda$ are odd.  Due to the $\mathbb{Z}_2$ symmetry of the background, the even and odd linearized fluctuations cannot mix.  Here we are interested in the resolution mode $\delta \lambda$, so we will focus on the mixing among the odd fluctuations.

The linearized equations are
\es{LinearizedEquations}{
{e^{{1 \over 2} w} \over r^3} \left({r^3 g \over e^{{1\over 2} w}} \delta \lambda' \right)'
+  e^{2 \eta - {4 \over 3} \chi + w} \Phi' \left[ \Phi' \delta \lambda
- \delta \tilde \Phi_R' \right]
+ {4 \over L^2} e^{-6 \eta - {8\over 3} \chi} (3 e^{5 \eta} - 2) \delta \lambda &= 0 \ , \\
{g \over r^3 e^{{1 \over 2} w}} (r^3 e^{-4 \eta + {8 \over 3} \chi+ {1 \over 2} w} \delta \Phi_R')'
- {16 \over L^2} e^{-4 \eta - 4 \chi } (\delta \Phi_R + \delta \tilde \Phi_R) &=0 \ , \\
{g \over r^3 e^{{1 \over 2} w}} (r^3 e^{2 \eta - {4\over 3} \chi+ {1 \over 2} w}  \delta \tilde \Phi_R')'
- 2 g e^{2 \eta - {4 \over 3} \chi} \delta \lambda' \Phi'
- {8 \over L^2} e^{-4 \eta - 4 \chi}  (\delta \Phi_R + \delta \tilde \Phi_R) &=0 \ ,
}
where $\chi$, $\eta$, $\Phi$, $g$, and $w$ are evaluated at their background values given in \eqref{PhieomSimp} and \eqref{GotLeading} for the zero-temperature extremal solution.  Let's focus on this extremal solution and find a series expansion in $r-1$ (we set $L = r_h=1$).  This calculation is similar to that of the non-analytic contributions to the background given at the end of section~\ref{sec:zeroT}.  Since \eqref{LinearizedEquations} is a system of three second order differential equations, there are six linearly independent solutions whose leading behaviors are of the form
\es{SolutionForm}{
\delta \Phi_R &= e^{{\alpha \over \tilde r}}
\tilde r^{b_1} \left(\delta \Phi_R^{(0)} + \delta \Phi_R^{(1)} \tilde r + \dotsb \right) \ , \\
\delta \tilde \Phi_R &= e^{{\alpha \over \tilde r}}
\tilde r^{b_2} \left(\delta \tilde \Phi_R^{(0)} + \delta \tilde \Phi_R^{(1)} \tilde r + \dotsb \right) \ , \\
\delta \lambda &= e^{{\alpha \over \tilde r}} e^{{5 \over 72 \tilde r}}
\tilde r^{b_3 - {21 \over 20}} \left(\delta \lambda^{(0)} + \delta \lambda^{(1)} \tilde r + \dotsb \right) \ ,
}
with $\tilde r \equiv r -1$ as in \eqref{Defs}.  The coefficients $\alpha$ and $b_i$ are given in table~\ref{PertCoefs}.
All six solutions satisfy the $U(1)_R$ charge conservation condition \eqref{ConservedRCharge} at the linearized level.  Of the modes that do not grow with $y$ (I, III, IV, and V), only mode III requires a non-vanishing $Q_R$, while the others satisfy the conservation equation with $Q_R = 0$.
\TABULAR {c||c|c|c|c|c}
{
Solution & $\alpha$ & $b_1$ & $b_2$ & $b_3$ & $\Delta_{\rm IR}$ \\
\hline \hline
I & $-{5 \over 36}$ & ${83\over 30}$ & $b_1 + 1$ & $b_1 + 1$ & $2$, source  \\
II & ${5 \over 72}$ & $-{163 \over 60}$ & $b_1 + 1$ & $b_1 + 1$ & $2$, VEV   \\
III & $-{5 \over 72}$ & ${21\over 20}$ & $b_1$ & $b_1$ & $0$, VEV\\
IV & $0$ & $0$ & $b_1$ & $b_1$ & $0$, source  \\
V & $-{5 ( 1 + \sqrt{5}) \over 144} $ & ${63 - 7 \sqrt{5} \over 120}$ & $b_1$ & $b_1$ & ${1 + \sqrt{5} \over 2}$, source  \\
VI & $-{5 ( 1 - \sqrt{5}) \over 144}$ & ${63 + 7 \sqrt{5} \over 120}$ & $b_1$ & $b_1$ & ${1 + \sqrt{5} \over 2}$, VEV}
{The coefficients of the perturbative expansion \eqref{SolutionForm} and the IR dimensions of the corresponding operators. The solution IV is in fact an exact pure gauge mode for which $\Phi_R = - \tilde \Phi_R = \text{const}$. \label{PertCoefs}}
The crucial fact is that $\alpha$ is real for all six solutions, so there are no oscillatory solutions in the near-$AdS_2$ region.  The absence of oscillatory solutions means that the black 3-branes with baryonic charge constructed in the previous sections are likely to be stable with respect to the perturbations \eqref{LinearizedEquations}.

We discussed in section~\ref{sec:zeroT} how near the extremal horizon, the geometry is $AdS_2 \times \mathbb{R}^3 \times T^{1, 1}$ up to slowly varying logarithmic factors.  We can thus ask what the effective dimensions of the operators dual to the modes given in table~\ref{PertCoefs} are.  Changing variables to the $AdS_2$ coordinate $y$ defined in \eqref{newCoord}, we see that $\delta \lambda$ behaves for the six solutions as $y^{-1}$, $y^2$, $y^{0}$, $y$, $y^{{1-\sqrt{5} \over 2}}$, and $y^{{1 + \sqrt{5} \over 2}}$, respectively.

Using \eqref{sourcevev}, the dimensions $\Delta_{\mathrm{IR}}$ corresponding to various perturbations are given in the last column of table~\ref{PertCoefs}.  Since solution IV is exact, has $\lambda \equiv 0$,
and is pure gauge, we suspect that the $\Delta_{\mathrm{IR}} = 0$ modes we are seeing correspond to the conserved charge operator in the effective quantum mechanics. This is consistent with the fact that mode III, which produces a VEV of this operator, is seen to correspond to non-vanishing $Q_R$. In this paper we only study solutions with vanishing R-charge, so the charged modes with $\Delta_{\mathrm{IR}}=0$ are not allowed. The remaining dimensions we find, $2$ and ${1 + \sqrt{5} \over 2}$, correspond to irrelevant operators from the point of view of the IR near-$AdS_2$ theory. The sources for such operators correspond to modes that fall off near the horizon as $y^{-1}$ or $y^{{1 - \sqrt{5} \over 2}}$. Since the operators are irrelevant, we expect that inducing them in the IR theory will not destroy the near-conformal IR solution we find.


\section{Discussion}
\label{sec:discussion}

In this paper we initiated studies of black hole solutions charged under baryonic symmetries. Such solutions are asymptotic to $AdS\times Y$, and the baryonic $U(1)_B$ symmetries appear due to the non-trivial topology of the Einstein space $Y$. We discussed the type IIB example $AdS_5\times T^{1,1}$ in some detail, and have also set up a consistent truncation for M-theory on $AdS_4\times Q^{1,1,1}$. Perhaps our most surprising finding is that the type IIB charged 3-brane solution develops, in the zero-temperature limit, a novel kind of near-horizon region, which is a warped product $AdS_2 \times {\mathbb R}^3\times T^{1,1}$ with warp factors that are logarithmic in the AdS radius. This supergravity solution is smooth because the logarithms decrease the curvature of the solution; in fact, all curvatures approach zero at the horizon. In this sense this solution is reminiscent of the UV region of another solution based on the conifold, with a topologically non-trivial $3$-form flux turned on \cite{Klebanov:2000nc,Klebanov:2000hb}. That warped deformed conifold solution was supersymmetric and automatically stable. In the present case, where the only non-vanishing supergravity fields are the metric and the self-dual $5$-form, the solution does not seem to preserve any supersymmetry, and its stability is a serious issue. We carried out some highly non-trivial stability checks for our solution.

We have shown that the simplest objects charged under the baryonic $U(1)_B$, namely the wrapped D3-branes, do not condense.
This still leaves the possibility that one of the neutral fields might cause an instability. Our solution preserves a certain $\mathbb{Z}_2$ symmetry and we have checked stability with respect to one of the modes odd under the $\mathbb{Z}_2$. This well-known mode, dual to the operator ${\rm Tr} (A_i\bar A_i- B_j\bar B_j)$, turns on the difference of the sizes of the two 2-spheres that is present in a small resolution of the conifold, and also mixes with the $U(1)_R$ gauge field. We leave further studies of stability for future work. We also note that we have encountered difficulties in extending our numerical solution all the way to zero temperature. At the lowest temperature we have been able to reach numerically, $w_h/2\approx 0.85$, which means that the near-$AdS_2$ throat is only beginning to develop.  It would be interesting to construct the full numerical $T=0$ solution that matches onto the near-horizon form that we found analytically. One of our goals is to study the fermion modes in this background, following the interesting results of \cite{Lee:2008xf,Liu:2009dm, Cubrovic:2009ye,Faulkner:2009wj}. There are various extensions of this work related to finding other black brane solutions with baryonic charges. We hope to present these results in the future.

\section*{Note Added}

After this paper was completed, we learned of the interesting work \cite{Hartnoll:2009ns} where another potential instability of charged brane backgrounds was suggested. Such an instability, called 
the ``Fermi seasickness'' in \cite{Hartnoll:2009ns}, is caused by nucleation of a space-time filling D-brane
towards the AdS boundary (for earlier discussions of similar
D-brane instabilities see \cite{Yamada:2008em, Evans:2002fk}).\footnote{We thank Eva Silverstein for suggesting to us that our background may suffer from this instability.} 
In the dual gauge theory, this corresponds to an instability with respect to the Coulomb branch, where
certain mesonic operators develop vacuum expectation values.
In the background we have studied, this instability can be seen from computing the potential for probe D3-branes filling the $(t, \vec{x})$ directions. Numerical computations show that for temperatures greater than about $0.2 \mu$ the charged black branes constructed in this paper are stable with respect to such D3-brane nucleation, while for lower temperatures they become metastable.
For any nonzero temperature the D3-brane is attracted near the horizon, which means that there exists
a potential barrier that for large $N$ prevents brane nucleation to AdS infinity.



\section*{Acknowledgments}

We would like to thank S.~Gubser, J.~Maldacena, A.~Nellore, J.~Polchinski, K.~Schalm, E.~Silverstein, D.~Son, E.~Witten, and A.~Yarom for useful discussions.
This work was supported in part by the US NSF  under Grant Nos.
PHY-0551164, PHY-0756966, and PHY-0844827. C.P.H. and I.R.K. would like to thank the Kavli Institute for Theoretical Physics for hospitality during the AdS/CMT workshop, which was stimulating for this project.


\appendix
\section{Small Charge Limit}
\label{sec:smallcharge}

Approximate solutions can be found in the small $Q$ limit.  Defining the dimensionless parameter $\tilde Q \equiv Q L / r_h^3$, the small $Q$ expansion takes the form
\es{SmallQAnsatz}{
w &= \tilde Q^2 \delta w^{(2)} + \tilde Q^4 \delta w^{(4)} + \dotsb \ , \\
g &=\frac{ r^2}{L^2} \left(1  - {r_h^4 \over r^4} \right)+ Q^2 \delta g^{(2)} + Q^4 \delta g^{(4)} + \dotsb \ , \\
\Phi &= {Q \over 2} \left({1 \over r_h^2} - {1 \over r^2}\right) + \tilde Q^3 \delta\Phi^{(3)} + \dotsb \ , \\
\eta &= \tilde Q^2 \delta \eta^{(2)} + \tilde Q^4 \delta \eta^{(4)} + \dotsb \ , \\
\chi &= \tilde Q^2 \delta \chi^{(2)} + \tilde Q^4 \delta \chi^{(4)} + \dotsb \ ,
}
where the starting point of the expansion obtained by setting $Q = 0$ corresponds to the AdS-Schwarzschild solution.  The small $Q$ approximation \eqref{SmallQAnsatz} can be alternatively thought of as a large temperature expansion.

To second order in $Q$, defining $\rho \equiv r/r_h$, one finds the solution
\es{Gotp2}{
\delta w^{(2)} &= 0  \ , \qquad
\delta g^{(2)} = {1 \over 12 r^4} \left(1 - \rho^2 \right) \ , \\
\delta \eta^{(2)} &= \frac{1}{20} \left(
\frac{ 1 - 4 \rho^4}{2 \rho^2 } -  Q_{-{3 \over 2}} \left( 1 - 2 \rho^4 \right)
\right) \ , \\
\delta \chi^{(2)} &= - \frac{1}{40} \left( {1 + 3  \rho^2 - 6 \rho^4 \over   \rho^2} + {3  (1 - 2 \rho^4) } \log {\rho^2 \over 1 + \rho^2} \right) \ .
}
This solution obeys the boundary conditions \eqref{BdySeries} at the conformal boundary and is also regular at the black hole horizon.

The thermodynamic quantities \eqref{GotThermo} become in this limit
\es{ThermoSmallQ}{
\rho &= {Q \over 2 \kappa_5^2 L^2 } \ , \qquad
s = {2 \pi r_h^3 \over \kappa_5^2 L^3 } \ , \qquad
\epsilon = {r_h^4 \over 8 \kappa_5^2 L^5} \left(12 + \tilde Q^2 \right)  \ , \\
\mu &= {Q \over 2 L r_h^2} \ , \qquad
T = {r_h \over 24 \pi L^2} \left( 24 - \tilde Q^2 \right) \, .
}
One can check explicitly that the relation \eqref{EpsilonVar} is satisfied.  It is also useful to note that the values $\eta_h$ and $\chi_h$ of $\eta$ and $\chi$ at the horizon are
\es{etachiHor}{
\eta_h = {\pi - 3 \over 40}\tilde Q^2 \ , \qquad
\chi_h = -{\log 8 - 2 \over 40} \tilde Q^2 \ .
}


\section{A Consistent Truncation for Baryonic 2-Branes in $AdS_4 \times Q^{1, 1, 1}$}
\label{sec:Qbranes}

The consistent truncation of $11$d SUGRA that contains one of the two $U(1)_B$ gauge fields can be obtained as follows.  The eleven-dimensional metric is
\es{11dMetric}{
ds^2 = e^{-7\chi/2} ds_M^2 + 4 L^2 e^\chi \Bigg[
{e^{\eta + \lambda} \over 8} \left(d\theta_1^2 + \sin^2 \theta_1 d\phi_1^2
+ d\theta_2^2 + \sin^2 \theta_2 d\phi_2^2 \right) \\
+ {e^{\eta - 2 \lambda} \over 8} \left(d\theta_3^2 + \sin^2 \theta_3 d\phi_3^2 \right)
+ {e^{-6 \eta} \over 16} \left(d\psi + \sum_{i = 1}^3 \cos \theta_i d\phi_i \right)^2\Bigg] \ .
}
Here, $ds_M^2$ is the line element on a non-compact four-dimensional space $M$.  When $\chi = \eta = \lambda = 0$, \eqref{11dMetric} describes $M \times Q^{1,1,1}$ \cite{Page:1984ae}.

The four-form $F_4$ can be constructed using the forms $\omega_2$ and $\omega_5$ defined as
\es{omega25Def}{
\omega_2 &\equiv {1\over 2} \left(\sin \theta_1 d\theta_1 \wedge d\phi_1
- \sin \theta_2 d\theta_2 \wedge d\phi_2 \right)  \ , \qquad  \\
\omega_5 &\equiv (d\psi + \cos \theta_1 d\phi_1
+ \cos \theta_2 d\phi_2) \wedge \omega_2 \wedge (\sin\theta_3 d\theta_3 \wedge d\phi_3) \ .
}
The four-form $F_4$ is then
\es{GotF4}{
F_4 = {3 \over L} e^{-{21\over 2}\chi} \vol_M + {L^2 \over \sqrt{2}}
e^{2 \eta + 2 \lambda - {3\over 2} \chi} (*_M F) \wedge \omega_2 \ .
}
Its dual, $F_7 = *F_4$, is given by
\es{GotF7}{
F_7 = {3 L^6 \over 8} \omega_2 \wedge \omega_5
+ {L^5 \over 4 \sqrt{2}} F \wedge \omega_5 \ .
}

This ansatz preserves the $\mathbb{Z}_2$ space-time inversion symmetry where $x^\mu\rightarrow - x^\mu$ is accompanied by the interchange of two 2-spheres, $(\theta_1,\phi_1)\leftrightarrow (\theta_2,\phi_2)$. This symmetry requires the metrics of these 2-spheres to be the same, thus eliminating an additional function.

Equations \eqref{11dMetric} and \eqref{GotF4} give a consistent truncation of eleven-dimensional SUGRA only when $F \wedge F = 0$.  The effective four-dimensional lagrangian is
\es{Lag4d}{
{\cal L}_{\rm eff} = R - {1 \over 4} e^{2 \eta + 2 \lambda - {3 \over 2} \chi} F_{\mu\nu}^2
- {63 \over 8} (\partial_\mu\chi)^2 - {21 \over 2} (\partial_\mu \eta)^2
- 3 (\partial_\mu \lambda)^2 - V(\eta, \chi, \lambda) \ ,
}
where
\es{GotV4d}{
V(\eta, \chi, \lambda) = {e^{-8 \eta - 2 \lambda - {21 \over 2} \chi} \over 2 L^2}
\left[ 9 e^{8 \eta + 2 \lambda} + 2 e^{6 \chi} + e^{6 (\lambda + \chi)}
-16\, e^{7 \eta + \lambda + 6 \chi} - 8 e^{7 \eta + 4 \lambda + 6 \chi}
\right] \ .
}

One can again look for extrema of \eqref{Lag4d} of the form \eqref{5dansatz}, except now there are only two coordinates along the brane, $dx^1$ and  $dx^2$. The equations of motion resulting from ${\mathcal L}_{\rm eff}$ are
\begin {subequations}
\label {4deoms}
\begin {align}
\eta'' + \eta' \left(\frac 2r + \frac {g'} g + \frac {21r} 4 \eta'^2 + \frac {63r} {16} \chi'^2 + \frac {3r}2 \lambda'^2\right) + \frac {\Phi'^2} {21g} e^{w + 2\eta - \frac 32 \chi + 2\lambda} - \frac 1 {21 g} \frac {\partial V} {\partial \eta} &= 0 \label {4deometa} \ , \\
\chi'' + \chi' \left(\frac 2r + \frac {g'} g + \frac {21r} 4 \eta'^2 + \frac {63r} {16} \chi'^2 + \frac {3r}2 \lambda'^2\right) - \frac {\Phi'^2} {21g} e^{w + 2\eta - \frac 32 \chi + 2\lambda} - \frac 4 {63g} \frac {\partial V} {\partial \chi} &= 0 \label {4deomchi} \ , \\
\lambda'' + \lambda' \left(\frac 2r + \frac {g'} g + \frac {21r} 4 \eta'^2 + \frac {63r} {16} \chi'^2 + \frac {3r}2 \lambda'^2\right) + \frac {\Phi'^2} {6g} e^{w + 2\eta - \frac 32 \chi + 2\lambda} - \frac 1 {6g} \frac {\partial V} {\partial \lambda} &= 0 \label {4deomlambda} \ , \\
g' + g \left(\frac 1r + \frac {21r} 4 \eta'^2 + \frac {63r} {16} \chi'^2 + \frac {3r} 2 \lambda'^2\right) + \frac {r \Phi'^2} 4 e^{w + 2\eta - \frac 32 \chi + 2\lambda} + \frac r2 V &= 0 \label {4deomg} \ , \\
\Phi'' + \Phi' \left(\frac 2r + \frac 12 w' + 2 \eta' - \frac 32 \chi' + 2 \lambda'\right) &= 0 \label {4deomphi}\ , \\
w' + 2 r \left(\frac {21}4 \eta'^2 + \frac {63} {16} \chi'^2 + \frac 32\lambda'^2\right) &= 0 \ .\label {4deomw}
\end {align}
Again, we can immediately integrate the $\Phi$ equation, yielding
\es {4deomphibetter} {
\Phi' &= \frac {Q} {r^2} e^{-\frac 12 w - 2 \eta + \frac 32 \chi - 2\lambda}\ ,
}
where $Q$ is an integration constant related to the charge of the black hole.
\end {subequations}
Plugging \eqref{4deomphibetter} into equations~\eqref {4deometa}--\eqref {4deomg}, we again eliminate all dependence on $w$,
\begin {equation}
\begin {aligned}
\eta'' + \eta' \left(\frac 2r + \frac {g'} g + \frac {21r} 4 \eta'^2 + \frac {63r} {16} \chi'^2 + \frac {3r}2 \lambda'^2\right) + \frac {Q^2} {21 r^4 g} e^{-2\eta + \frac 32 \chi - 2\lambda} - \frac 1 {21 g} \frac {\partial V} {\partial \eta} &= 0 
\ , \\
\chi'' + \chi' \left(\frac 2r + \frac {g'} g + \frac {21r} 4 \eta'^2 + \frac {63r} {16} \chi'^2 + \frac {3r}2 \lambda'^2\right) - \frac {Q^2} {21 r^4 g} e^{-2\eta + \frac 32 \chi - 2\lambda} - \frac 4 {63g} \frac {\partial V} {\partial \chi} &= 0 
\ , \\
\lambda'' + \lambda' \left(\frac 2r + \frac {g'} g + \frac {21r} 4 \eta'^2 + \frac {63r} {16} \chi'^2 + \frac {3r}2 \lambda'^2\right) + \frac {Q^2} {6 r^4 g} e^{-2\eta + \frac 32 \chi - 2\lambda} - \frac 1 {6g} \frac {\partial V} {\partial \lambda} &= 0 
\ ,  \\
g' + g \left(\frac 1r + \frac {21r} 4 \eta'^2 + \frac {63r} {16} \chi'^2 + \frac {3r} 2 \lambda'^2\right) + \frac{Q^2}{4 r^3} e^{-2\eta + \frac 32 \chi - 2\lambda} + \frac r2 V &= 0   \ .
\end {aligned}
\end {equation}
We hope to report on solutions of these equations in a later publication.

\bibliographystyle{JHEP}
\bibliography{baryon}

\end{document}